\documentclass[sigconf]{acmart}

\AtBeginDocument{%
  \providecommand\BibTeX{{%
    \normalfont B\kern-0.5em{\scshape i\kern-0.25em b}\kern-0.8em\TeX}}}

\setcopyright{acmcopyright}
\copyrightyear{2018}
\acmYear{2018}
\acmDOI{XXXXXXX.XXXXXXX}

\acmConference[Conference acronym 'XX]{Make sure to enter the correct
  conference title from your rights confirmation emai}{June 03--05,
  2018}{Woodstock, NY}
\acmBooktitle{Woodstock '18: ACM Symposium on Neural Gaze Detection,
 June 03--05, 2018, Woodstock, NY} 
\acmPrice{15.00}
\acmISBN{978-1-4503-XXXX-X/18/06}

\begin{document}

\title{Demonstrating Eye Movement Biometrics in Virtual Reality}

\author{Dillon Lohr}
\orcid{0000-0002-8088-9270}
\affiliation{%
  \institution{Texas State University}
  \streetaddress{601 University Drive}
  \city{San Marcos}
  \state{Texas}
  \country{USA}
  \postcode{78666-4684}
}
\email{djl70@txstate.edu}

\author{Saide Johnson}
\orcid{}
\affiliation{%
  \institution{Texas State University}
  \streetaddress{601 University Drive}
  \city{San Marcos}
  \state{Texas}
  \country{USA}
  \postcode{78666-4684}
}
\email{s_j242@txstate.edu}

\author{Samantha Aziz}
\orcid{0000-0002-7656-2662}
\affiliation{%
  \institution{Texas State University}
  \streetaddress{601 University Drive}
  \city{San Marcos}
  \state{Texas}
  \country{USA}
  \postcode{78666-4684}
}
\email{sda69@txstate.edu}

\author{Oleg Komogortsev}
\orcid{0000-0001-7890-8842}
\affiliation{%
  \institution{Texas State University}
  \streetaddress{601 University Drive}
  \city{San Marcos}
  \state{Texas}
  \country{USA}
  \postcode{78666-4684}
}
\email{ok11@txstate.edu}

\renewcommand{\shortauthors}{Lohr, et al.}

\begin{abstract}
Thanks to the eye-tracking sensors that are embedded in emerging consumer devices like the Vive Pro Eye, we demonstrate that it is feasible to deliver user authentication via eye movement biometrics.
\end{abstract}




\maketitle

\section{Introduction}
There is an emerging use of eye-tracking sensors in virtual reality devices like the Vive Pro Eye and in augmented reality devices like the HoloLens~2.
The inclusion of these eye-tracking sensors is motivated in part to enable foveated rendering~\cite{guenter2012foveated}, which can make higher-resolution displays viable in tethered devices and can offer significant power savings in untethered devices.
As such, these devices already include the hardware necessary for performing user authentication via eye movement biometrics.
We believe that eye movement biometrics could become the leading biometric authentication technique in these devices, similar to how fingerprint and facial recognition have become ubiquitous in smartphones.

Eye movement biometrics is a behavioral biometric modality that has been thoroughly studied since its introduction in 2004~\cite{Kasprowski2004}.
Most studies focus on high quality eye-tracking signals~\cite{george2016score,Friedman2017,makowski2021deepeyedentificationlive,lohr2022ekyt}.
The current state-of-the-art model~\cite{lohr2022ekyt} is able to approach a level of authentication accuracy suitable for real-world use using these high quality signals.
However, modern virtual- and augmented-reality headsets exhibit a much lower level of signal quality than the datasets used for such studies.

Prior live eye movement biometrics frameworks~\cite{holland2014framework,lohr2018implementation} also use lower levels of signal quality but often do not report results.
The present study can be differentiated by its use of a more modern model and by the fact that we report performance measures.
Additionally, the model we employ is pre-trained using artificially degraded eye-tracking signals from a different eye tracker.
The model did not see data from the eye-tracking device we use in this study---the Vive Pro Eye---prior to our final evaluation.

\section{Training the Model}
\begin{figure*}
    \centering
    \includegraphics[width=\linewidth]{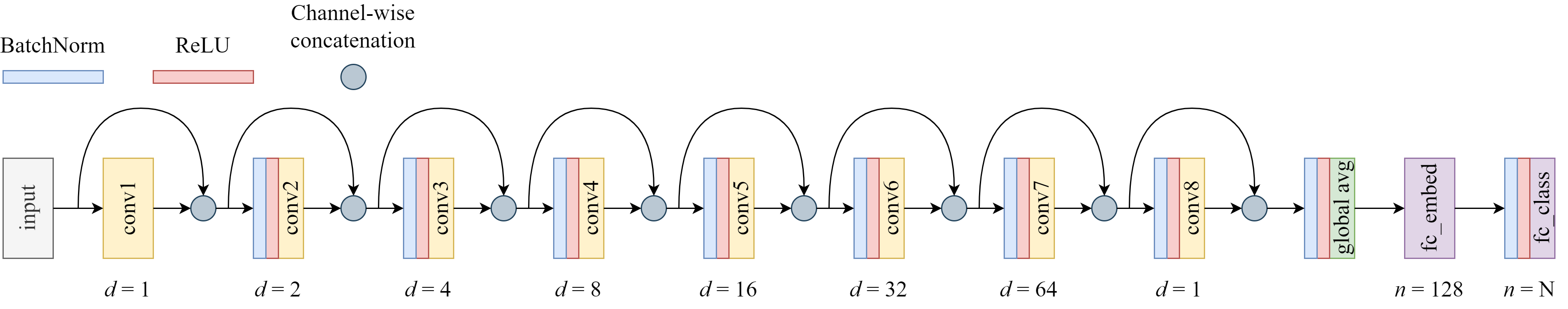}
    \caption{The network architecture from Eye Know You Too. Figure from~\cite{lohr2022ekyt}.}
    \label{fig:network}
\end{figure*}

We employ the state-of-the-art Eye Know You Too architecture~\cite{lohr2022ekyt} which is a DenseNet-based convolutional neural network.
The network is visualized in Figure~\ref{fig:network}.
The model is trained using the GazeBase dataset~\cite{griffith2021gazebase} which comprises recordings from 322~college-aged participants.
Each participant was present for up to 18~recording sessions, and each session consisted of 7~different eye-tracking tasks: horizontal saccades~(HSS), random saccades~(RAN), reading~(TEX), an interactive ball-popping game~(BLG), two video-viewing tasks~(VD1 and VD2), and a fixation task~(FXS).

We follow the training methodology described in the Eye Know You Too manuscript~\cite{lohr2022ekyt} with a couple differences to better fit our purposes.
The first difference is that we artificially degrade the eye-tracking signals by downsampling from 1000~Hz to 125~Hz and by adding white (Gaussian) noise with mean~0 and standard deviation~0.1 to the training data after performing z-score normalization.
We chose a sampling rate of 125~Hz to closely match the 120~Hz sampling rate of the Vive Pro Eye, and the addition of spatial noise was intended to improve performance on the lower signal quality of the Vive Pro Eye.
The second difference is that we do not train an ensemble of models, but rather we train only the one model that uses the first data fold as the validation set.
The use of only one model instead of the ensemble of 4~models was mostly intended to minimize the computational burden of the model at inference time, but using the ensemble would likely have led to superior performance.

\section{Live Biometrics Demo}
The live demo is built with Unity and supports both enrollment and verification.
The user enters their name (or the name of a different enrolled user if acting as an impostor) in a text field, performs the manufacturer-provided calibration procedure, and then performs a 9-second eye-tracking task similar to the RAN task from GazeBase.
During the demo task, a small, black stimulus appears over a light-gray background and jumps between the positions of a standard 9-point grid with a 1-second period between jumps.

The full 9-second recording is fed at once into the model to produce a 128-dimensional embedding that acts as a user's biometric template.
During enrollment, the produced embedding is saved to a database for later verification attempts.
During verification, the produced embedding is compared against the enrolled embedding.
It takes less than a second to produce an embedding from the eye-tracking signal.

\section{User Study}
To evaluate our model's performance, we collected data from 5~participants (3 male, 2 female) with normal or corrected-to-normal vision.
Each participant (labeled A--E) was recorded twice using the Vive Pro Eye.
The first recording (e.g., A1) is used for enrollment, and the second recording (e.g., A2) is used for verification against each participant's enrolled template (A1--E1).

Our measured similarity scores are reported in Table~\ref{tab:similarities}.
\begin{table}
    \centering
    \caption{Similarity scores between the enrollment and verification embeddings collected from each participant. Bolded scores represent falsely accepted/rejected verification attempts at a threshold of 0.8.}
    \label{tab:similarities}
    \begin{tabular}{c|ccccc}
        \toprule
        Enroll \textbackslash{} Verify & A2 & B2 & C2 & D2 & E2 \\
        \midrule
        A1 & 0.8119 & 0.7014 & 0.7155 & 0.6596 & 0.7473 \\
        B1 & 0.7814 & 0.8551 & 0.6858 & 0.7664 & 0.7377 \\
        C1 & \textbf{0.8034} & 0.7441 & \textbf{0.7987} & 0.6468 & \textbf{0.8169} \\
        D1 & \textbf{0.8440} & \textbf{0.8546} & 0.7580 & 0.8169 & 0.7613 \\
        E1 & 0.7718 & 0.7974 & 0.7160 & 0.6904 & 0.8494 \\
        \bottomrule
    \end{tabular}
\end{table}
We achieve an equal error rate~(EER) of 20\% (1-in-5 false rejection rate and 4-in-20 false acceptance rate) with a similarity score threshold of~0.8.
It is important to keep in mind that our model was not trained on data from the Vive Pro Eye, yet it still performed better than chance.

\section{Conclusion}
We demonstrated eye movement biometrics in virtual reality using the Vive Pro Eye.
We achieved 20\% EER on a dataset of 5~participants using a deep learning model pre-trained on a different dataset with no fine-tuning.
Once larger datasets in virtual reality become available (e.g.,~\cite{lohr2020vr}), better performance could likely be achieved.

\begin{acks}
This material is based upon work supported by the National Science Foundation Graduate Research Fellowship under Grant No. DGE-1144466 and DGE-1840989.
Any opinions, findings, and conclusions or recommendations expressed in this material are those of the author(s) and do not necessarily reflect the views of the National Science Foundation.
\end{acks}

\bibliographystyle{ACM-Reference-Format}
\bibliography{main}


\begin{thebibliography}{10}


\ifx \showCODEN    \undefined \def \showCODEN     #1{\unskip}     \fi
\ifx \showDOI      \undefined \def \showDOI       #1{#1}\fi
\ifx \showISBNx    \undefined \def \showISBNx     #1{\unskip}     \fi
\ifx \showISBNxiii \undefined \def \showISBNxiii  #1{\unskip}     \fi
\ifx \showISSN     \undefined \def \showISSN      #1{\unskip}     \fi
\ifx \showLCCN     \undefined \def \showLCCN      #1{\unskip}     \fi
\ifx \shownote     \undefined \def \shownote      #1{#1}          \fi
\ifx \showarticletitle \undefined \def \showarticletitle #1{#1}   \fi
\ifx \showURL      \undefined \def \showURL       {\relax}        \fi
\providecommand\bibfield[2]{#2}
\providecommand\bibinfo[2]{#2}
\providecommand\natexlab[1]{#1}
\providecommand\showeprint[2][]{arXiv:#2}

\bibitem[Friedman et~al\mbox{.}(2017)]%
        {Friedman2017}
\bibfield{author}{\bibinfo{person}{Lee Friedman}, \bibinfo{person}{Mark~S.
  Nixon}, {and} \bibinfo{person}{Oleg~V. Komogortsev}.}
  \bibinfo{year}{2017}\natexlab{}.
\newblock \showarticletitle{Method to assess the temporal persistence of
  potential biometric features: Application to oculomotor, gait, face and brain
  structure databases}.
\newblock \bibinfo{journal}{\emph{PLOS ONE}} \bibinfo{volume}{12},
  \bibinfo{number}{6} (\bibinfo{date}{06} \bibinfo{year}{2017}),
  \bibinfo{pages}{1--42}.
\newblock
\urldef\tempurl%
\url{https://doi.org/10.1371/journal.pone.0178501}
\showDOI{\tempurl}


\bibitem[George and Routray(2016)]%
        {george2016score}
\bibfield{author}{\bibinfo{person}{Anjith George} {and}
  \bibinfo{person}{Aurobinda Routray}.} \bibinfo{year}{2016}\natexlab{}.
\newblock \showarticletitle{A Score Level Fusion Method for Eye Movement
  Biometrics}.
\newblock \bibinfo{journal}{\emph{Pattern Recogn. Lett.}} \bibinfo{volume}{82},
  \bibinfo{number}{P2} (\bibinfo{date}{oct} \bibinfo{year}{2016}),
  \bibinfo{pages}{207–215}.
\newblock
\urldef\tempurl%
\url{https://doi.org/10.1016/j.patrec.2015.11.020}
\showDOI{\tempurl}


\bibitem[Griffith et~al\mbox{.}(2021)]%
        {griffith2021gazebase}
\bibfield{author}{\bibinfo{person}{Henry Griffith}, \bibinfo{person}{Dillon
  Lohr}, \bibinfo{person}{Evgeny Abdulin}, {and} \bibinfo{person}{Oleg
  Komogortsev}.} \bibinfo{year}{2021}\natexlab{}.
\newblock \showarticletitle{GazeBase, a large-scale, multi-stimulus,
  longitudinal eye movement dataset}.
\newblock \bibinfo{journal}{\emph{Scientific Data}} \bibinfo{volume}{8},
  \bibinfo{number}{1} (\bibinfo{date}{16 Jul} \bibinfo{year}{2021}),
  \bibinfo{pages}{184}.
\newblock
\urldef\tempurl%
\url{https://doi.org/10.1038/s41597-021-00959-y}
\showDOI{\tempurl}


\bibitem[Guenter et~al\mbox{.}(2012)]%
        {guenter2012foveated}
\bibfield{author}{\bibinfo{person}{Brian Guenter}, \bibinfo{person}{Mark
  Finch}, \bibinfo{person}{Steven Drucker}, \bibinfo{person}{Desney Tan}, {and}
  \bibinfo{person}{John Snyder}.} \bibinfo{year}{2012}\natexlab{}.
\newblock \showarticletitle{Foveated 3D Graphics}.
\newblock \bibinfo{journal}{\emph{ACM Trans. Graph.}} \bibinfo{volume}{31},
  \bibinfo{number}{6}, Article \bibinfo{articleno}{164} (\bibinfo{date}{nov}
  \bibinfo{year}{2012}), \bibinfo{numpages}{10}~pages.
\newblock
\urldef\tempurl%
\url{https://doi.org/10.1145/2366145.2366183}
\showDOI{\tempurl}


\bibitem[Holland and Komogortsev(2014)]%
        {holland2014framework}
\bibfield{author}{\bibinfo{person}{Corey~D. Holland} {and}
  \bibinfo{person}{Oleg~V. Komogortsev}.} \bibinfo{year}{2014}\natexlab{}.
\newblock \showarticletitle{Software Framework for an Ocular Biometric System}.
  In \bibinfo{booktitle}{\emph{Proceedings of the Symposium on Eye Tracking
  Research and Applications}}. \bibinfo{pages}{365–366}.
\newblock
\urldef\tempurl%
\url{https://doi.org/10.1145/2578153.2582174}
\showDOI{\tempurl}


\bibitem[Kasprowski and Ober(2004)]%
        {Kasprowski2004}
\bibfield{author}{\bibinfo{person}{Pawe{\l} Kasprowski} {and}
  \bibinfo{person}{J{\'{o}}zef Ober}.} \bibinfo{year}{2004}\natexlab{}.
\newblock \showarticletitle{{Eye movements in biometrics}}.
\newblock \bibinfo{journal}{\emph{Lecture Notes in Computer Science}}
  \bibinfo{volume}{3087} (\bibinfo{year}{2004}), \bibinfo{pages}{248--258}.
\newblock
\urldef\tempurl%
\url{https://doi.org/10.1007/978-3-540-25976-3_23}
\showDOI{\tempurl}


\bibitem[Lohr et~al\mbox{.}(2018)]%
        {lohr2018implementation}
\bibfield{author}{\bibinfo{person}{Dillon Lohr}, \bibinfo{person}{Samuel-Hunter
  Berndt}, {and} \bibinfo{person}{Oleg Komogortsev}.}
  \bibinfo{year}{2018}\natexlab{}.
\newblock \showarticletitle{An Implementation of Eye Movement-Driven Biometrics
  in Virtual Reality}. In \bibinfo{booktitle}{\emph{Proceedings of the 2018 ACM
  Symposium on Eye Tracking Research \& Applications}}. Article
  \bibinfo{articleno}{98}, \bibinfo{numpages}{3}~pages.
\newblock
\urldef\tempurl%
\url{https://doi.org/10.1145/3204493.3208333}
\showDOI{\tempurl}


\bibitem[Lohr and Komogortsev(2022)]%
        {lohr2022ekyt}
\bibfield{author}{\bibinfo{person}{Dillon Lohr} {and} \bibinfo{person}{Oleg~V
  Komogortsev}.} \bibinfo{year}{2022}\natexlab{}.
\newblock \bibinfo{title}{Eye Know You Too: A {DenseNet} Architecture for
  End-to-end Eye Movement Biometrics}.
\newblock
\newblock
\showeprint[arXiv]{2201.02110}~[cs.CV]


\bibitem[Lohr et~al\mbox{.}(2020)]%
        {lohr2020vr}
\bibfield{author}{\bibinfo{person}{Dillon~J Lohr}, \bibinfo{person}{Samantha
  Aziz}, {and} \bibinfo{person}{Oleg Komogortsev}.}
  \bibinfo{year}{2020}\natexlab{}.
\newblock \showarticletitle{Eye Movement Biometrics Using a New Dataset
  Collected in Virtual Reality}. In \bibinfo{booktitle}{\emph{ACM Symposium on
  Eye Tracking Research and Applications}}. Article \bibinfo{articleno}{40},
  \bibinfo{numpages}{3}~pages.
\newblock
\urldef\tempurl%
\url{https://doi.org/10.1145/3379157.3391420}
\showDOI{\tempurl}


\bibitem[Makowski et~al\mbox{.}(2021)]%
        {makowski2021deepeyedentificationlive}
\bibfield{author}{\bibinfo{person}{Silvia Makowski}, \bibinfo{person}{Paul
  Prasse}, \bibinfo{person}{David~R. Reich}, \bibinfo{person}{Daniel
  Krakowczyk}, \bibinfo{person}{Lena~A. Jäger}, {and} \bibinfo{person}{Tobias
  Scheffer}.} \bibinfo{year}{2021}\natexlab{}.
\newblock \showarticletitle{DeepEyedentificationLive: Oculomotoric Biometric
  Identification and Presentation-Attack Detection Using Deep Neural Networks}.
\newblock \bibinfo{journal}{\emph{IEEE Transactions on Biometrics, Behavior,
  and Identity Science}} \bibinfo{volume}{3}, \bibinfo{number}{4}
  (\bibinfo{year}{2021}), \bibinfo{pages}{506--518}.
\newblock
\urldef\tempurl%
\url{https://doi.org/10.1109/TBIOM.2021.3116875}
\showDOI{\tempurl}


\end{thebibliography}

\end{document}